# Self-similar transformations of lattice-Ising models at critical temperatures

You-gang Feng

College of Science, Guizhou University, Huaxi, Guiyang, 550025 China

E-mail: ygfeng45@yahoo.com.cn

**Abstract.** We classify geometric blocks that serve as spin carriers into simple blocks and compound blocks by their topologic connectivity, define their fractal dimensions and describe the relevant transformations. By the hierarchical property of transformations and a block-spin scaling law we obtain a relation between the block spin and its carrier's fractal dimension. By mapping we set up a block-spin Gaussian model and get a formula connecting the critical point and the minimal fractal dimension of the carrier, which guarantees the uniqueness of a fixed point corresponding to the critical point, changing the complicated calculation of critical point into the simple one of the minimal fractal dimension. The numerical results of critical points with high accuracy for five conventional lattice-Ising models prove our method very effective and may be suitable to all lattice-Ising models. The origin of fluctuations in structure at critical temperature is discussed. Our method not only explains the problems met in the renormalization-group theory, but also provides a useful tool for deep investigation of the critical behaviour.

## 1. Introduction

To explore a new effective method we should sum up those methods that have been proved effective. Series expansion is a good approximate method. By some introduced physical conditions, it can be applied to calculate the critical points [1-3]. There are two kinds: the high-temperature expansion and the low-temperature expansion. The convergence of a series makes us believe that the method may give us a result infinitely close to an exact solution. Apart from the results, we get nothing about the critical phenomena, especially the fluctuation structure. In the most occasions the results are used to test and verify the effectiveness of other methods as mentioned in [4-5]. Kramers and Wannier developed duality transformation method [6]. By topologic structures of the geometric lattices, a trigonal lattice and a hexagonal lattice are dual to one another; the dual lattice of a tetragonal lattice is still a tetragonal lattice. By Euler theorem one can obtain an equivalent relation between the partition functions of the dual lattice systems. If there is only one singular point in the relation equation, the point is just known as the critical point of the original lattice. They obtained some exact solutions for these dual lattice systems. However, we cannot carry the method on in *3*-dimensional space for an exact calculation. Onsager connected the singularity of free energy with the critical point satisfying many experimental facts [7]. Using matrix method he calculated the partition function of square planar lattice and found an exact formula of the free energy, such that he got the exact critical point. The difficulty has never been overcome when his method is used in the calculation of partition function for a *3*-dimensional model. Mean field approximation is a phenomenological one because its some order parameters must be determined by experimental data. It merely gives us more or less rough results; much effective information about the system structures is wiped out as introducing the mean field [8].

In the 1960's Fisher and Widow put forward the initial ideas of scaling laws to study critical phenomena, the ideas helped Kadanoff mold his scaling law and a new concept, Kadanoff block, which is a lattice-spin set keeping all symmetries of the original lattice and behaving as magnetic order [9-11]. Under his law the partition function of lattice spins is replaced by the partition function of block spins, and the block spin system has the same critical properties as the original ones. By his suggestion there is only one kind of block with the same side. The side is governed by one's ability to compute. The greater the ability is, the longer the side will be.

Adopting Kadanoff scaling law, Wilson set forward the renormalization-group theory to study Ising models [12]. Here, we only discuss it, for more easy comparison, in the real space since our theory is suitable to the space. To obtain a critical point one should finish seven steps below: (1) determine Kadanoff block of side $n$; (2) set an effect Hamiltonian of the block spins; (3) formulate the block-spin variant; (4) define a partial trace; (5) calculate the trace; (6) find the relevant renormalization transformations; (7) determine the fixed point of the transformations. The theory regards the fixed point as the critical point. By some approximate procedure such as the coarse graining or the decimation, the original lattice spin system is changed into the block-spin system. The block Hamiltonian is similar to the lattice's one. At this stage the first transformation is made, the transformation of lattice spins into block spins is designated as $(R_n)^1$. On the same thinking, we can construct a new lager block spin with these block spins formed by $(R_n)^1$ after rescaling on the Kadanoff scaling law; this is called the second transformation denoted by $(R_n)^2$. Clearly, such a kind of transformation can be exerted further. There are also hierarchical transformations $(R_n)^3$, $(R_n)^4$, …. The number of times for the transformations will be infinity because there are infinite lattices in the model. Adding in an identical transformation $(R_n)^0$, we get a set of transformations: $(R_n)^0$, $(R_n)^1$, $(R_n)^2$, $(R_n)^3$, $(R_n)^4$, …, of the set consists a hemigroup $\{(R_n)^r, r = 0,1,2,3,\cdots\}$; the prefix "hemi" denotes there is no inverse transformation. The key point is that when a system approaches to the critical point the correlation length becomes infinity and its scaling transformations will not change, such that one can obtain the critical point by finding the fixed point of transformations rather than computing the partition function.

However, with the repeating cumulate expansion during the calculation of partial trace much more parameters are introduced. The more terms remain the more parameters arise, such that the final calculation cannot go on. Reviewing his theory in a lecture [13], Wilson said: "A serious problem with the renormalization-group transformations is that there is no guarantee that they will exhibit fixed points, …, iteration of a critical point, does not lead to a fixed point", "There is no known principle for avoiding this possibility, and as Kadanoff has shown using his decimation procedure, a simple approximation to a transformation can misleadingly give a fixed point when the full transformation cannot". The fact reminds us that the theory has not set a principle to stipulate the block size at the critical point, a critical point corresponds to a fixed point may be not a full condition, but a necessary one.

Not having accurate solution urges authors to research for other methods. Z. D. Zhang suggested a distinction method [4]. He thinks that the difficulty in the calculation of critical point of a *3*-dimensional model is that the solution may be linked to the fourth curled –up dimension. He proposed two conjectures subject to a boundary condition, he then got some solutions for the model. Wu F Y and his colleagues doubt the conjectures and comment on his work [5]. His two conjectures are purely mathematic. In addition, the added boundary condition may change the topologic structure of the original system [14]. Being response to their comment, Zhang admitted that there are some open questions out of his conjecture, which need more research [15]. We think that the solution structure is concerned in the lattice one, hence the most important thing is to research the topologic structure of lattice system itself. People didn't realize there is a relationship between the critical phenomena and the fractal theory until Wilson published his lecture [13], he said: "There is a murky connection between scaling ideas in critical phenomena and Mandelbrot's 'fractals' theory – a description of scaling of irregular geometrical structures, …" (Wilson p595-596). The renormalization-group transformation is a kind of self-similar transformation relating to the fractals. His ideas stimulated some authors to study the fractal dimensions of the blocks by Monte Carlo method [16-17]. It seems that there is blindness in choice of block sides. That there exists a fixed point of the transformation is a thing, a critical point relates to a fixed point is another thing. The results of Monte Carlo method show that different block sizes have different fractal dimensions. Now that the critical point is unique, what size leads to the critical point? A restrictive relationship between the singularity of free energy and the fixed point was neglected. We should have a full condition to guarantee that a special fixed point will uniquely determine the critical point.

In this paper we try to explain what causes the difficulties met by Wilson's theory, and to explore a united effective method to study the critical phenomena for 2-and-3-dimensional lattice-Ising models. We inquire into the fractal structures of blocks, combining which with the critical characteristic we try to find a fixed point linking to a critical point. In fact, an Ising model consists of a geometric lattice system and spins. For example, a cubic lattice spin system involves a cubic geometric lattice system and spins; a lattice spin means a spin residing in a lattice site. A block spin is a block in which each lattice site has a spin and these lattice spin correlation makes the block act as magnetic order. Clearly, a lattice and a block are the carriers of a lattice spin and a block spin, respectively. There is certain relationship between those spins and their carriers (the duality transformations also do relate to the carrier's ones). The primary thing is that the topologic connectivity of a block determines whether a block spin can become magnetic order. In addition, when a renormalization group is used for a lattice spin system a self-similar transformation of its geometric lattice system is exerted at the same time; therefore both transformations are one-to-one. In view of isomorphism the group can be also used to describe the carrier's transformations. In section 2 we analyze at first the connectivity of blocks, classify them into two kinds by their topologic properties. We then define the fractal dimensions for blocks. Further, we put forward a block-spin scaling law for the spin self-transformations, finding a relation between a block spin and a fractal dimension of its carrier, which helps us avoid complicated calculating a block spin. By mapping, we set up a block-spin Gaussian model that can be solved exactly, and get a formula connecting a critical point with the minimal fractal dimension of the carrier. In section 3, at first we calculate the critical points for five conventional lattice-Ising models, the results with high accuracy test and verify our method very effective. By our theory we then discuss the fluctuation origin in structure.

## 2. Theory
### 2.1 Block classification

Only those fractal dimensions of the blocks corresponding to magnetic order are physically meaningful. A carrier of a disordered block is equal to an empty set without any dimension. The lattice spin system is order if and only if its carrier is a simply connected space [14]. In the mathematic sense, the simplest lattices are called simplex, that is a simply connected space and can shrink to a lattice. Trigonal lattices and tetrahedron lattices are the simplexes [18]. Other lattices are called complexes, each of them is not simply connected and cannot shrink to a lattice. By these principles, we classify the blocks into two kinds. The first is called simple blocks, involving the trigonal lattice blocks and the tetrahedron lattice blocks. Others such as the square planar lattice blocks and the cubic lattice blocks belong to the second kind, called compound blocks. By the topologic rule a compound block ought to be decomposed into $k$ parts, each part is simply connected, called sub-block. The thermodynamic equilibrium and the transformation uniformity require the sub-blocks identical. Such a division is purely mathematic, and the block is regarded as a sum set of the $k$ sub-blocks, each sub-block is called a sub-space of the block. If we put one spin at each lattice site of a sub-block, we then form a sub-block spin. Clearly, a sub-block is a carrier of a sub-block spin. The $k$ sub-block spins correlate to each other to make the compound block magnetic order, such a process is purely physical. Generally, a compound block spin is said to be order when its carrier is a product space of its $k$ sub-spaces, since the product space is simply connected [18], while these $k$ sub-block spins act as a whole. By differential geometry the inside space of a sub-block can be considered as a vector space of $D$ dimensions. An ordered compound block spin is a product space of its $k$ vector sub-spaces, called a $D^k$-dimensional vector space [19].

For a compound block spin system there are two types of block spins: a *D*-type spin represents a sub-block spin; a $D^k$-type spin relates to that the *k* *D*-type spins in a block act as a whole with carrier's dimensions $D^k$. There are only *D*-type spins in a simple block spin system. The denotation links a block spin or a sub-block spin with its carrier's dimensions.

*2.2 Transformation groups and fractal dimensions of carriers*

Kadanoff scaling idea is to replace a lattice spin and a lattice spin system by a block spin and a block spin system, respectively. The replacement is accompanied by the self-similar transformations of relevant carriers. An original lattice is called a zero-order lattice. We suppose that there are infinite zero-order lattices on the zero-th hierarchy. The element $(R_n)^1$ of the group $\{(R_n)^r, 0,1,2,3,\cdots\}$ forms a first-order block on the first hierarchy by *P* zero-order lattices. Element $(R_n)^2$ forms a second-order block on the second hierarchy by *P* first-order blocks, containing $P^2$ zero-order lattices. Generally, the element $(R_n)^r$, $r > 1$, forms an r-order block on the r-th hierarchy by *P* (r-1)-order blocks, containing $P^r$ zero-order lattices. Since a carrier is a set of discrete zero-order lattices, box-counting dimension definition is feasible to the fractal structures of blocks [20]. The smallest number of sets to cover the zero-order lattices in an r-order block on the r-th hierarchy is $P^r$, and the diameter at most for every set is $1/n^r$. We will see from equation (1) that the transformations manifest a scaling law, by which the *n* is the block side. For the infinitely hierarchical transformation, $r \to +\infty$, the diameter tends to vanish, $(1/n^r) \to 0$. The fractal dimension of an r-order block on the r-th hierarchy is

$$D = -\lim_{r \to \infty} \frac{\ln P^r}{\ln(1/n^r)} = \frac{\ln P}{\ln n} \quad (1)$$

where the symbol "ln" is a natural logarithm, and the value of *n* should guarantee the dimension reasonable. The transformation is strictly self-similar, so the result $D = (\ln P)/(\ln n)$ also is the definition of similarity dimension [20]. Equation (1) states that an r-order block has a constant fractal dimension for any possible number *r*. The scaling law is that each r-order block contains *P* r-order lattices, which are originally the (r-1)-order blocks on the (r-1)-th hierarchy. As an r-order lattice its inner space should be omitted on the r-th hierarchy. The distance between nearest neighbor r-order lattices is consistently assigned as a unit length, and the block side is *n*. A special case is that a first-order block on the first hierarchy contains *P* first-order lattices, which are just the zero-order lattices on the zero-th hierarchy, where there is no block. By equation (1) an r-order block may be regarded as a *D*-dimensional hypercube of side *n*. We may understand the transformation from two hands through the equation: a definite block will determine a special dimension, or a certain dimension will require a particular block side.

Using equation (1), we can directly define the fractal dimension for a simple block. A trigonal block of side $n$ with lattices $P = (n+1)(n+2)/2$ is shown in figure 1. The fractal dimension $D_{tr}$ is

$$D_{tr} = \frac{\ln[(n+1)(n+2)/2]}{\ln n} \qquad (2)$$

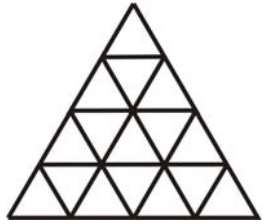

**Figure 1**. A trigonal-lattice block of side $n = 4$

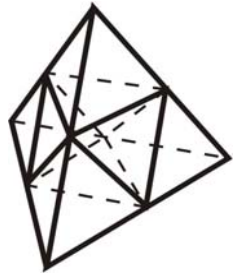

**Figure 2**. A tetrahedron-lattice block of side $n = 2$

With the same reason, the fractal dimension $D_{te}$ of a tetrahedron block is

$$D_{te} = \frac{\ln[(n+1)(n+2)(n+3)/6]}{\ln n} \qquad (3)$$

The block is plotted in figure 2.

For a kind of simple blocks with side $n$ there is a group $\{(R_n)^r, r = 0,1,2,3,\cdots\}$, different sides have different attendant groups. The self-similar transformation is a kind of contraction map, there is a fixed point in the transformation due to the contraction-mapping theorem [21]. The question is that for all possible sides which is responsible for the critical point? We should find a relation connecting a particular side $n^*$ and a critical point $K_c$. A special group $\{(R_{n^*})^r, r = 0,1,2,3,\cdots\}$ with $n^*$ relates to a particular transformation that will determine $K_c$. From equation (1) we know that the $n^*$ will result in a unique fractal dimension that may determine the $K_c$ by a certain restrictive condition.

We shouldn't directly define a fractal dimension of a compound block by equation (1) in that on the one hand since it is a complex and cannot become order; on the other hand the direct definition of a compound block by equation (1) will make its fractal dimension larger than $d$, its embedding dimension. The formation of an ordered compound block spin can be thought of as involving two steps: Some lattice spins at first construct a sub-block spin called $D$-type spin; every $k$ $D$-type spins then proceed by their correlation to form a $D^k$-type spin that represents the $k$ $D$-type spins act as a whole, such that the transformation from lattice spins into ordered compound spins is self-similar. Every $D^k$-type spin can shrink to a lattice spin due to its simply connectivity, a larger $D^k$-type spin will form through the previous steps. The two types of transformations are necessary, otherwise the self-similarity will be broken. When we consider the $D^k$-type spin interaction the $D$-type spins are involved in the $D^k$-type spins, the k $D$-type spins in a block act as a whole such that there doesn't exist an interaction between an independent $D^k$-type spin and a $D$-type spin. Therefore, there are two independent sub-systems in a compound block spin system: the $D$-type spin system relating to $D$-type spin interaction and the $D^k$-type spin system to $D^k$-type spin interaction. Whether the $D$-type spin formation and the $D^k$-type spin formation or their relevant interactions and correlations are all in all from the interaction and long-range correlation of lattice spins. By these reasons and from the self-similar transformation viewpoint we suppose there are two independent groups: one is group $\{(R_{n,s})^r, r = 0,1,2,3,\cdots\}$ for the $D$-type spins; another is group $\{(R_{n,o})^r, r = 0,1,2,3,\cdots\}$ for the $D^k$-type spins. The element $(R_{n,s})^j$, $1 < j < +\infty$, changes some sub-blocks formed by $(R_{n,s})^{j-1}$ into new larger sub-blocks; and the element $(R_{n,o})^j$, $1 < j < +\infty$, changes the ordered compound blocks formed by $(R_{n,o})^{j-1}$ into new larger ordered compound blocks; and $(R_{n,s})^j$ and $(R_{n,o})^j$ correspond to the same hierarchy. The two groups are only equivalent descriptions, but the $D$-type spins and the $D^k$-type spins exist really. A group $\{(R_{n^*,s})^r, r = 0,1,2,3,\cdots\}$ with a particular side $n^*$ results in a critical point $K_{c1}$. A group $\{(R_{n^*,o})^r, r = 0,1,2,3,\cdots\}$ with the same side $n^*$ determines a critical point $K_{c2}$.

For the equal description of the group $\{(R_{n,s})^r, r = 0,1,2,3,\cdots\}$ we can define the fractal dimensions of sub-blocks by equation (1). The decomposition of a complex into simplexes is called triangulation in the mathematic sense [18]. We notice that in a square planar lattice spin system there is no trigonal lattice that is a simplex. If it existed there would be the next nearest neighbor interaction. As a result, a simplex with the physical meaning is obliged to be a rectangular lattice. The reason is also suitable to a cubic lattice; its simplex is a cuboid. Let there be $k$ sub-blocks in a square planar block, a case $k = 2$ is illustrated in figure 3. If $k = 3$, an additional sub-block would be in between the two sub-blocks, it would not transform in the same way as the other: If the short sides of two neighbor sub-blocks changed into semi-infinity, its short side would have to keep limit rather than semi-infinity as restrain on it placed by the other sub-blocks. Such non-uniform transformation will break down the similarity, so that the case $k = 3$ is impossible. If $k = 4$, there are four identical small squares, each of them is a complex as well. The cases $k > 4$ are similar to the case $k = 3$ or to the case $k = 4$. Therefore, the case $k = 2$ is a unique choice. The division of a compound block with lattices $(n+1)^2$ into $k = 2$ sub-blocks doesn't change the relative positions of lattices, so the diameter of a covering open ball is still $1/n$. By equation (1), we get the fractal dimension of a sub-block:

$$D_{sq} = \frac{\ln[(n+1)^2/2]}{\ln n} \tag{4}$$

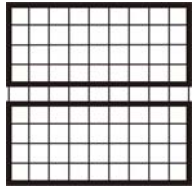

**Figure 3**. A square planar lattice block of side $n=9$ containing $k=2$ sub-blocks

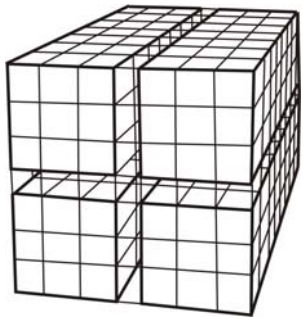

**Figure 4**. A cubic lattice block of side $n=7$ containing $k=4$ sub-blocks

Similarly, the fractal dimension of a sub-block of the cubic block is

$$D_{cu} = \frac{\ln[(n+1)^3/4]}{\ln n} \tag{5}$$

As shown in figure 4, a cube subdivides into $k = 4$ cuboids. If $k = 2$, a sub-block would have two identical square lattices on its 2-dimensional sections. The square lattice has the same side as the cube side. If we define the fractal dimension of the square lattice its dimension will be greater than the dimension of section as being its carrier, thereby the case $k=2$ impossible. So does the case $k=3$. For $k=8$ there will be 8 identical cubic blocks with smaller sides, each of them still is a complex. Other cases are similar to the case $k=2$, or $k=3$, or $k=8$, respectively. This implies that a cubic block only contains four identical sub-blocks. Thus, the fractal dimensions of carrier for an ordered square block spin ($k=2$) are $D_{sq}^2$, and a carrier of an ordered cubic block spin ($k=4$) is of $D_{cu}^4$ dimensions. For example, in the cubic block spin system there are $D$-type spins and $D^4$-type spins.

*2.3 Block-spin scaling law*

The original lattice spins are on the zero-th hierarchy, where doesn't exist any block spin. Under the operation of $(R_n)^r, 1 < r < +\infty$, some $(r-1)$-order block spins form an $r$−order block spin on the $r$−th hierarchy. At the same time these $(r-1)$-order block spins become the inner lattice spins of the $r$−order block spin. As a lattice spin its inside space is indistinguishable. The scaling law for spins is: When an object serves as a block spin, its spin is $S$ with coordination number $Z_1$ and coupling constant $J_B$; when it serves as an inner lattice spin, its spin is $s$ with coupling constant $j_L$ and coordination number *2D*, where the *D* is the fractal dimension of the block, which is regarded as a hypercube (the inside space of a block is *D* dimensions, its inner lattice sites are arranged with the regularity of *D*-dimensional hypercube, such that their coordination number is *2D*. For example, 4 for the square lattice; 6 for the cubic lattice). For the same object, its total coupling energy should keep conservation under the transformation, leading to $Z_1 J_B S^2 = 2Dj_L s^2$. There should be only one independent variant in an identity, such that we let $J_B$ equal $j_L$ (the equivalency is only available to a certain block spin, different block spins with different sides have different equal relations). Notice that $s^2 = 1$, we then get a relation between a block spin and its carrier's fractal dimension:

$$Z_1 S^2 = 2D \qquad (6.1)$$

Equation (6.1) is applicable to all *D*-type spins. By the equivalent description of the ordered compound block spins, the scaling law is the same as saying that the element $(R_{n,o})^j$ changes the $(j-1)$−order ordered compound block spins into the $j$−order ordered compound block spins on the $j-$th hierarchy, and these $(j-1)-$order ordered block spins originally being on the $(j-1)-$th hierarchy become now the inner lattice spins of the $j-$order block spins. Therefore, with the same reason as equation (6.1), for a $D^k$-type spin $S$ with coordination number $Z_2$ we have

$$Z_2 S^2 = 2D^k \tag{6.2}$$

Although the carrier's transformation and the block spin transformation are one-to-one, the singularity of free energy is determined by the block spin system other than the carrier's. A restrictive relation between the singularity of free energy and the fractal dimension should be found to guarantee that a special fractal dimension uniquely determines the critical point.

*2.4 Block-spin Gaussian model*

At first, we don't distinguish temporarily a block spin from a sub-block spin, and call each of them a block spin. On the $r-$th hierarchy infinite block spins relate to a group element $(R_n)^r$. For a definite $r$, the total number of the $r-$th hierarchy are great enough but infinity for all possible elements $(R_n)^r$ with different finite values of $n$. Each value of $n$ corresponds to a statistical system, where infinite block spins are independent of one another when the temperature is higher than $T_c$. Certainly, the number of the statistical systems is so great that we can make use of the statistical law to describe them. In the thermodynamic equilibrium state there are the fluctuations of block side $n$, obeying the Gaussian distribution. We now set up a new lattice spin system, keeping the original symmetry. In the new system there are infinite new lattice spins, all of them have the same magnitude $S$, being in either spin-up state or spin-down state. The spin values can change and correspond to the previous block spin values, and the change obeys the Gaussian law. We call the new lattice spin system the block-spin Gaussian model similar to the lattice-spin Gaussian model, known as spherical one exactly solvable [22]. Different $D$-type spin systems or $D^k$-type spin systems have their own Gaussian models. We take the $D$-type spin system of trigonal lattices as an example to explain its block-spin Gaussian model in detail. Let the partition function of the original trigonal lattice spin system be $Q_{tr}$, the partition function of its Gaussian model be $Q_{G,tr}$. The function $Q_{tr}$ is given by $Q_{tr} = Q_{G,tr} + Q_{s,tr}$. The function $Q_{s,tr}$ is a supplement to the $Q_{G,tr}$, involving all other possible spin configurations apart from the model's. According to the Ergodic hypothesis the model's configurations may occur. We think such description of $Q_{tr}$ may be either completely correct or good approximation. The spin configurations of $Q_{s,tr}$ include both block spins and lattice spins, it will have not be concerned in the self-similar transformation at $T_c$. Therefore, the critical singularity of free energy of the Gaussian model will determine the one of original system. For the convenience, in the following the new lattice spin and the $D$-type spin have the same meaning for $Q_{G,tr}$. It is

$$Q_{G,tr} = \int_{-\infty}^{+\infty} \cdots \int_{-\infty}^{+\infty} \prod_{j=1}^{N_c} \exp[K \sum_{(i,j)} S_i S_j - (2<S_{tr}^2>)^{-1} \sum_{j=1}^{N_c} S_j^2] \, dS_j \quad (7)$$

where $S_i$, $S_j = \pm S_{tr}$, $S_{tr}$ is the $D$-type spin magnitude, $<S_{tr}^2>$ is its mean square, $N_c$ is the total number of the solid-state physical primitive cell for the new system, and the sum $\sum_{(i,j)}$ is over all nearest neighbors, $K = J_{tr}/(k_B T)$, $J_{tr}$ the coupling constant, $k_B$ Boltzmann constant, $T$ temperature. For simplicity, we extend $S_i$ and $S_j$ to infinity: $-\infty < S_i, S_j < +\infty$, and allow the spins to take on continuous values. Such procedure is due to we only focus on the singularity of the free energy rather than the function value. We introduce Fourier transformations:

$$S_i = (1/\Omega) \sum_q S_q \exp i\, \boldsymbol{q} \cdot \boldsymbol{r}_i, \quad S_q = V \sum_{i=1}^{N_c} S_i \exp{-i\, \boldsymbol{q} \cdot \boldsymbol{r}_i} \quad (8)$$

where $V$ is a cell volume and $\Omega = N_c V$, $\boldsymbol{r}_i$ the position vector of $S_i$, $S_q$ is a spin with the value $q$, $\boldsymbol{q}$ the reciprocal lattice vector of the new system reflecting the new lattice spin state, the sum $\sum_q$ is over the range of $q$ in the first Brillouin-zone involving all spin states. The final form of $Q_{G,tr}$ takes

$$Q_{G,tr} = \prod_q \{(2\pi\Omega V)[(<S_{tr}^2>)^{-1} - K(\boldsymbol{q})]^{-1}\}^{1/2} \quad (9)$$

$$K(\boldsymbol{q}) = K \sum_\delta \exp{-i\, \boldsymbol{q} \cdot \boldsymbol{\delta}_{ij}} \quad (10)$$

The denotation $\boldsymbol{\delta}_{ij}$ represents a vector from the lattice $i$ to its nearest neighbor $j$. The model free energy is given by

$$F = -k_B T \ln Q_{G,tr} = (\frac{k_B T}{2})\{\sum_q \ln[(<S_{tr}^2>)^{-1} - K(\boldsymbol{q})]\} + T \cdot const \quad (11)$$

The components of $\boldsymbol{q}$ are in the bellow ranges: $-4\pi/(3a) \leq q_x < 4\pi/(3a)$, $-2\pi/(a\sqrt{3}) \leq q_y < 2\pi/(a\sqrt{3})$, the lattice constant is $a$. From equation (11), the singularity of free energy occurs when the $K(\boldsymbol{q})$ equals $(<S_{tr}^2>)^{-1}$. Whenever the temperature $T$ is higher than $T_c$ the $K(\boldsymbol{q})$ is always less than $(<S_{tr}^2>)^{-1}$ to guarantee $F$ meaningful. Thus, the approach of $K(\boldsymbol{q})$ to $(<S_{tr}^2>)^{-1}$ leads to the maximum of $K(\boldsymbol{q})$, inversely the minimum of $<S_{tr}^2>$. For equation (10) the end-point coordinates of the six $(Z=6)$ lattice vectors associated with $\delta_{ij}$ near the origin are $(\pm a, 0)$, $[\pm a/2, (a\sqrt{3})/2]$, and $[\pm a/2, -(a\sqrt{3})/2]$. Inserting these into equation (10), the $K(\boldsymbol{q})$ reaches its maximum at $\boldsymbol{q}=0$. So we have $K(0) = 6K_c = (<S_{tr}^2>_{min})^{-1}$, where $K_c = J_{tr}/(k_B T_c)$. To guarantee $K(0)$ to be the maximum the term $<S_{tr}^2>_{min}$ should be the minimum, so that $<S_{tr}^2>_{min} = S_{tr,min}^2$, where $S_{tr,min}$ is the minimum spin. Thus, we get $K_c = (6S_{tr,min}^2)^{-1}$, where the number 6 is just the coordination number of a $D$-type spin. The calculation signifies that the zero vector $\boldsymbol{q}=0$ leads to the $K_c$, so do other systems. With the same reason, the critical point formula of a system is

$$K_c = \frac{1}{ZS_{min}^2} \quad (12)$$

where $Z$ is the coordination number of a $D$-type spin, its minimum spin is $S_{min}$. For the simple block spin system, using equations (6.1) and (12), we have

$$K_c = \frac{1}{2D_{min}} \quad (13)$$

In a compound block spin system there are two Gaussian models, one is for the $D$-type spins, another the $D^k$-type spins; using equations (6.1), (6.2) and (12), we get

$$K_{c1} = \frac{1}{2D_{min}} \quad , \quad K_{c2} = \frac{1}{2D_{min}^2} \quad (14)$$

where the meanings of $K_{c1}$ and $K_{c2}$ are explained in the sub-section 2.2. The singularity of the system free energy will be determined by the singularities of the two models. For example, in the cubic lattice spin system the interaction between nearest neighbor $D$-type spins is along the directions parallel to the sub-block short sides, see figure 4. Infinite sub-blocks construct a 2-dimensional system because the symmetric centers of these sub-blocks are on the same plane. There are $N_G$ such planes parallel to each other, and on each plane there is a $D$-type spin system. There is not any interaction between the systems, since the direction of $D$-type spin interaction is parallel to the planes. Denote the partition function of Gaussian model of the 2-dimensional spin system by $Q_{G1}$, and the partition function of $N_G$ such models by $(Q_{G1})^{N_G}$ due to each model exists independently. Denote the partition function of Gaussian model of $D^4$-type spins ($k = 4$) by $Q_{G2}$, and the partition function of the original system by $Q_{cu}$. We get $Q_{cu} = (Q_{G1})^{N_G} \cdot Q_{G2} + Q_s$, the product form of $(Q_{G1})^{N_G} \cdot Q_{G2}$ indicates each model exists independently, and there is no interaction between them. The function $Q_s$ is a supplement to these models, like the $Q_{s,tr}$ in the trigonal lattice spin system. Let the free energies $F$, $F_1$ and $F_2$ relate to $Q_{cu}$, $Q_{G1}$ and $Q_{G2}$, respectively. With the same reason as equation (11), we have $F = N_G F_1 + F_2 + T_c \cdot Const$ at $T_c$. The function $Q_s$ vanishes at $T_c$ since it has no relation to the singularity of $F$. The reason is also suitable to other compound block spin systems.

A subsequent relation that a critical point of a compound block spin system is attributed to the critical-point sum of its sub-systems can be introduced by the follows. The concept of primitive cell in solid-state physics is greatly helpful for us to understand matter's properties. There is always one atom per primitive cell in a crystal with one type of atoms. If the primitive cell is a parallelepiped with atoms at each of the eight corners, each atom is shared among eight cells, so that the total number of atoms that can represent the cell to interact independently with another cell is one: $8 \times (1/8) = 1$. A similar case is met here. Refer to figure 3, the interaction of nearest neighbor sub-block spins is along the direction parallel to the sub-block short side; in the direction a block containing two sub-blocks coordinates to two blocks. Refer to figure 4, the interactions of sub-block spins are along the directions parallel to the sub-block short sides; in the directions one block having four sub-blocks coordinates to four blocks. The facts imply when two compound block spins interact each block spin only donates one $D$-type spin that is independently involved in the block-spin interaction in every related direction. Let the interacting operator for two compound block spins along the direction be $U$, the interacting operator for two $D$-type spins be $U_1$. We also should consider the $D^k$-type spin interaction: the operator for two $D^k$-type spins is $U_2$. The ground state with a compound block spin is denoted $\varphi$, the ground state with a $D$-type spin $\varphi_1$, and the ground state with a $D^k$-type spin $\varphi_2$. The two compound block spin interaction along the direction parallel to the sub-blocks short side involves the previous two types of interactions, so we have $U = U_1 + U_2$, and $\varphi = \varphi_1 + \varphi_2$. An ordered block spin is of simple connectivity, so we can use the position vector of its symmetric center to represent its site. Let $\boldsymbol{r}_1$ and $\boldsymbol{r}_2$ be the position vectors for the compound block spins $A$ and $B$, respectively. Their ground-state wave functions are $\varphi_A(\boldsymbol{r}_1) = \varphi_{A1}(\boldsymbol{r}_1) + \varphi_{A2}(\boldsymbol{r}_1)$ and $\varphi_B(\boldsymbol{r}_2) = \varphi_{B1}(\boldsymbol{r}_2) + \varphi_{B2}(\boldsymbol{r}_2)$, where $\varphi_{A1}$ and $\varphi_{B1}$ are similar to $\varphi_1$, $\varphi_{A2}$ and $\varphi_{B2}$ similar to $\varphi_2$. In fact, a coupling constant is the exchange integral of its relevant wave functions [23]. Thus, we have $J = \iint \varphi_A^*(\boldsymbol{r}_1)\varphi_B^*(\boldsymbol{r}_2)U\varphi_A(\boldsymbol{r}_2)\varphi_B(\boldsymbol{r}_1)\,\mathrm{d}v_1\,\mathrm{d}v_2$, $J_1 = \iint \varphi_{A1}^*(\boldsymbol{r}_1)\varphi_{B1}^*(\boldsymbol{r}_2)U_1\varphi_{A1}(\boldsymbol{r}_2)\varphi_{B1}(\boldsymbol{r}_1)\,\mathrm{d}v_1\,\mathrm{d}v_2$, where $\mathrm{d}v_1$ and $\mathrm{d}v_2$ are the volume elements, and $J_2 = \iint \varphi_{A2}^*(\boldsymbol{r}_1)\varphi_{B2}^*(\boldsymbol{r}_2)U_2\varphi_{A2}(\boldsymbol{r}_2)\varphi_{B2}(\boldsymbol{r}_1)\,\mathrm{d}v_1\,\mathrm{d}v_2$. In the expansion of the exchange integral for $\varphi_A$ and $\varphi_B$ there is not any coherent term of $\varphi_{A1}$ with $\varphi_{B2}$ or of $\varphi_{A2}$ with $\varphi_{B1}$ as the $D$-type spin and the $D^k$-type spin belong to two independent sub-systems, respectively. We then get

$$J = J_1 + J_2 \tag{15}$$

Clearly, these coupling constants relate to the fractal dimensions, since the $D$-type spin and the $D^k$-type spin rely on ones. For the same critical temperature $T_c$ their related critical points are expressed by $K_c = J/(k_B T_c)$, $K_{c1} = J_1/(k_B T_c)$ and $K_{c2} = J_2/(k_B T_c)$, respectively. Using equations (14)-(15), we have

$$K_c = K_{c1} + K_{c2} = \frac{1}{2D_{min}} + \frac{1}{2D_{min}^k} \tag{16}$$

where $K_c$ is the critical point of the compound block spin system, it is just the critical point of the original lattice spin system. Equation (16) justifies the conclusion that the singularity of the system free energy will be determined by the singularities of its sub-systems', and vice versa.

### 3. Discussion
*3.1 Fixed points of block sides*
The scale invariance itself means there exists a fixed point of the scaling variable, as which the block side acts in the carrier transformation. From equation (2) we get

$$f(n) = [(n+1)(n+2)/2]^{1/D_{tr}} = n \tag{17}$$

By equation (17), a certain $D_{tr}$ gives a special fixed-point equation $f(n) = n$; different $D_{tr}$ lead to different equations, each of which has a unique fixed point of the $n$ if $n$ is finite as the fixed-point theorem [21]. Among them an equation with the minimum $D_{tr,min}$ has a solution $n^*$ linking to the critical point. For convenience, we can directly use equation (2) with an extreme condition $dD_{tr}/dn=0$, and get

$$n^* = 14.4955 \quad , \quad D_{tr,min} = 1.8141 \tag{18}$$

The numerical calculation for those values of $n$, $n > n^*$ or $n < n^*$, confirms that the minimum $D_{tr,min}$ is unique to guarantee the uniqueness of the critical point. With the same reason, for the tetrahedron lattice by equation (3), we get

$$n^* = 8.7272 \quad , \quad D_{te,min} = 2.4547 \tag{19}$$

Similarly, for the sub-block of square planar lattice, by equation (4) we have

$$n^* = 7.8400 \quad , \quad D_{sq,min} = 1.7800 \tag{20}$$

For the sub-block of cubic lattice by equation (5) we obtain

$$n^* = 4.7491 \quad , \quad D_{cu,min} = 2.4781 \tag{21}$$

*3.2 Verification*
Inserting equation (18) to equation (13), we get a result for the trigonal lattice spin system $K_c = 0.2756$. Kramers and Wannier got $K_c = 0.2747$, [6]. For the square planar lattice spin system $(k = 2)$, using equations (20) and (16), we get $K_c = 0.4387$. Onsager obtained $K_c = 0.4407$, [7]. For the cubic lattice system $(k = 4)$, by equations (21) and (16), we have $K_c = 0.2150$. The result of the high-temperature expansion is $0.2217$, [1]; the low-temperature expansion [3], is $0.222$; the result of Kramers-Wannier approximation [24], is $0.2184$; the Monte Carlo method got $0.2217$ and $0.2216544$ respectively [25-26]; the result of combinatorial and topological approach is $0.2188$ [27]. Inserting equation (19) into equation (13), we get $K_c = 0.2037$ for the tetrahedron lattice spin system; the result has not been reported so far except ours. Perhaps, its coordination number $Z = 14$ is too large to overcome the difficulty in its calculation by other ways. There is a generally recognized rule for those lattice spin systems with the same dimensionality that the larger the coordination number, the smaller the critical point value [28]. Comparing it with the result of the cubic lattice spin system with coordination number $Z = 6$, by the rule we judge our result to be reliable. The discussion about the critical point with high accuracy and the fluctuations of a hexagonal lattice spin system is in Appendix A. As far as the accuracy of these data is concerned, our theory has been proved enough validity rather than fortuitous coincidence.

Parameter $K = J/(k_B T)$ has a fixed point $K_c$, to find $K_c$ is to find $n^*$ due to the $J$ at the critical temperature is determined by $n^*$ as seen previously in the wave-function exchange integral and equations (16) and (1). It is the regardless of both the topologic connectivity of blocks and the restrictive relation between the block-side fixed point and the singularity of free energy in the renormalization-group theory, as Wilson said, that there is no guarantee that the transformations will exhibit fixed points relevant to the critical points.

*3.3 Fluctuation origin*
The critical fluctuations result from the adjustment of inner structures. For the self-similar transformations the structure adjustment is just the block side one. We see that on the one hand the transformations only allow the blocks to take on integer sides keeping the self-similarity; on the other hand a critical point just requires a fractal side. The deviation in the side is just the fluctuation. Additionally, in the sense of physics the fractal side makes some lattice spins be outside blocks or sub-blocks, these lattice spins have the same ability to form block spins or sub-block spins as the inner lattice spins, such that the system adjusts the block side to make all lattice spins be involved in new block spins or sub-block spins. The fluctuation-dissipation theorem points out that the fluctuations will automatically disappear at the thermodynamic equilibrium without foreign field [29]. On this principle, the system accommodates the block side successively to the critical point, while the point acts as an attractive center, around it

Self-similar transformations of lattice-Ising models at critical temperatures the side adjustment goes on from time to time. In the process a variety of block spins with integer sides will arise. We can keep a system to be at the critical temperature, but we are not able to impose a transformation of fractal side on to the system, so the fluctuations will forever exist.

## 4. Conclusion

That the fixed point of the self-similar transformation of a lattice-spin system relates to the critical point of the system as a necessary condition is foreshowed by the renormalization-group theory; the minimal fractal dimension, as a full one, will guarantee the existence of such a fixed point is proved in this paper. Our theory, which emphasizes the topologic connectivity of blocks, in principle is suitable to all lattice-Ising models, and simplifies the complicated calculation of critical points with high accuracy, having special meaning for the 3d lattice system that relevant calculation usually is difficult and tedious. Its analyses on lattice system structures help us deeply understand structural characteristic of critical behaviour, providing us a useful tool to research further the inner structure of fluctuations at the critical temperature [30].

## Appendix A

Figure A1 illustrates a compound block containing six sub-blocks for a hexagonal lattice system, for simplicity, where a triangle represents a sub-block. The detail structure of a sub-block is illustrated in figure A2, where a cell is a minimal hexagon.

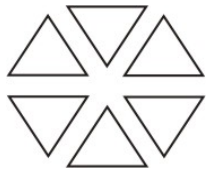

**Figure A1**. A hexagon-lattice block containing six sub-blocks

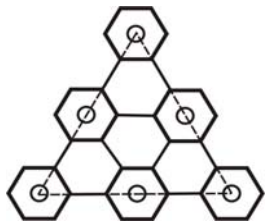

**Figure A2**. A sub-block of side $n=10$

In the figure a small circle denotes the cell center and all circles constitute an equilateral triangle. The total number of cells of six lattices increases in series of natural numbers with increasing the triangle side value. Thus, the total number of lattices in a sub-block is $P = 6 \cdot (1 + 2 + \cdots)$. Only those cells that lie on the triangle boundary contribute their partial girths to the girth of the sub-block. Let the girth of a cell be $L$, and $L = 6$, where the space of adjacent lattices is a unit length. A cell on a vertex angle of the triangle in figure A2 contributes five sixths of $L$ to the girth of sub-block, a cell not on the vertex one half of $L$ to the girth, a line-segment connecting two cells on the boundary one sixth of $L$. One third of the sub-block girth is the sub-block side, denoted by $n$. We then have $P = 6 \cdot [1 + 2 + \cdots + (n+2)/4]$, the fractal dimension of a sub-block is

$$D_{he} = \frac{\ln\{6 \cdot [1 + 2 + \cdots + (n+2)/4]\}}{\ln n} \qquad (A.1)$$

Where $n = 6, 10, 14, 18, 22, \ldots$ other values will change the sub-block boundary form to break down the self-similarity. The six $D$-type spins in a block spin shouldn't act as a whole, otherwise the blocks will shrink to lattices to form a trigonal lattice system dual to the hexagonal lattice system with other than the symmetry of the original system. To keep the original symmetry the $D$-type spins form two types of ordered units through correlations that are product spaces of the $D$-type spins. Every two $D$-type spins which centers lie on the same straight line and symmetry about the block center form the first type. There are three such units in a block as shown in figure A3, where a small circle represents a $D$-type spin, the line-segments represent that they act as a whole. Each unit is $D_{he}^2$ dimensions, called $D^2$-type spin. Every three $D$-type spins form another type of ordered unit; the symmetric centers of spins construct an equilateral triangle. There are two such units in a block shown in figure A4, and each unit is $D_{he}^3$ dimensions called $D^3$-type spin. The thermodynamical equilibrium and the transformation uniformity only allow the sub-block spins to form the two types of ordered units. The sub-blocks keep the original symmetry completely, they can construct new larger sub-blocks through the self-similar transformations. Six new sub-blocks form a new larger compound block, and the two types of ordered units of the new sub-blocks will appear in the new block. There is a group relating to the $D$-type spin transformation. However, the $D^2$-type spins and the $D^3$-type spins arise from the $D$-type spin correlations rather than the self-similar transformation. To describe their transformations in terms of self-similar transformations we can use the equivalent descriptions like the ones of other ordered compound block spins such as the square planar lattices. When two block spins interact along the direction of a straight line connecting two sub-block spins in a $D^2$-type spin shown in figure A3, each of them contributes one $D$-type spin, one $D^2$-type spin, and one $D^3$-type spin, corresponding to three Gaussian models, respectively. Therefore, the critical point of the original system is

$$K_c = \frac{1}{2D_{he,\min}} + \frac{1}{2D^2_{he,\min}} + \frac{1}{2D^3_{he,\min}} \tag{A.2}$$

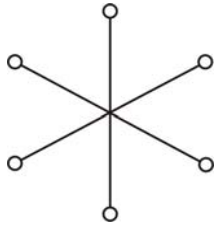

**Figure A3**. Three ordered units of two sub-blocks in a block

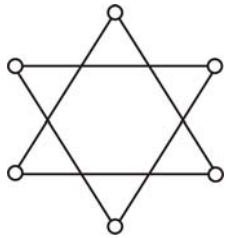

**Figure A4**. Two ordered units of three sub-blocks in a block

The numerical calculation of equation (A.1) yields

$$n^* = 14 \quad , \quad D_{he,\min} = 1.5514 \tag{A.3}$$

Inserting equation (A.3) into equation (A.2), we get $K_c = 0.6639$, $K_c = 0.6585$ is the result of the duality transformation method [6].

Referring figures A3 and A4, a $D^2$-type spin and a $D^3$-type spin share commonly a sub-block in a block, by this way the block becomes their partially connected space such that the system can avoid forming a trigonal lattice spin system. Without the two types of spins the system will not become magnetic order after infinite iteration. The sub-block spin formation cannot involve all lattice spins, referring to figure A2, there are some odd lattice spins outside the sub-block spin. Because of the Kadanoff scaling law the block spin system is similar to the lattice spin system, these odd lattice spins form certainly a disordered region to amount to an empty set. These lattice spins, however, have the same ability to form sub-block spins as the inner lattice spins. That a lattice spin lies inside a sub-block spin or lies outside the sub-block spin results in a deviation in spin state, it forces the system to adjusts the side to make all lattice spins be contained in new sub-block spins to eliminate the deviation. However, there are always lattice spins outside, whether what side the sub-block will have. Thus, the side adjustment never stop, this is just the fluctuation cause.